\documentclass[12pt,preprint]{aastex}
\usepackage{psfig}

\slugcomment{Accepted for publication in the Astrophysical Journal}

\shortauthors{Strickland \etal}
\shorttitle{The luminous X-ray source in NGC 3628 reappears}

\newcommand{\eg}{{\rm e.g.\ }}

\newcommand{\ie}{{\it i.e.\ }}

\newcommand{\etal}{{\rm et al.\thinspace}}

\newcommand{\cm}{{\rm\thinspace cm}}

\newcommand{\s}{{\rm\thinspace s}}

\newcommand{\erg}{{\rm\thinspace erg}}
\newcommand{\ps}{\hbox{\s$^{-1}\,$}}

\newcommand{\ergps}{\hbox{$\erg\s^{-1}\,$}}

\newcommand{\pcmsq}{\hbox{$\cm^{-2}\,$} }

\newcommand{\nH}{\hbox{$N_{\rm H}$}}

\newcommand{\pc}{{\rm\thinspace pc}}

\newcommand{\Msol}{\hbox{$\thinspace M_{\sun}$}}

\begin{document}

\title{Another intermediate mass black hole in a starburst galaxy?:
	The luminous X-ray source in NGC 3628 reappears}

\author{David K. Strickland\footnote{{\it Chandra} Fellow}, \, 
	 Edward J.M. Colbert and Timothy M. Heckman}
\affil{Department of Physics and Astronomy, The Johns Hopkins University,
	3400 North Charles Street, Baltimore, MD 21218}
\email{dks@pha.jhu.edu, colbert@pha.jhu.edu, heckman@pha.jhu.edu}

\author{Kimberly A. Weaver}
\affil{NASA/Goddard Space Flight Center, Code 662, Greenbelt, Maryland 20771}
\email{kweaver@cleo.gsfc.nasa.gov}

\author{Michael Dahlem}
\affil{European Southern Observatory, Castilla 19001, Santiago 19, Chile}
\email{mdahlem@eso.org}

\and
\author{Ian R. Stevens}
\affil{School of Physics \& Astronomy, The University of Birmingham,
	Edgbaston, Birmingham, B15 2TT, U.K.}
\email{irs@star.sr.bham.ac.uk}

\begin{abstract}
In a 52 ks-long {\it Chandra} ACIS-S observation of the nearby starburst
galaxy NGC 3628, obtained to study the starburst-driven outflow from
this galaxy, we have detected a very luminous ($L_{\rm X} \approx 1.1
\times 10^{40} \ergps$ in the 0.3 -- 8.0 keV energy band) point
source located at least
$20\arcsec$ ($\sim 970$ pc) from the
nucleus of the galaxy. 
No radio, optical or near-IR counterpart to this source has been found.

This is most probably the reappearance of the
strongly-variable
X-ray-luminous source discovered by \citet{dahlem95}, which faded
by a factor $\ga 27$ between December 1991 and March 1994 (at which point
it had faded below the detection limit in a {\it ROSAT} HRI observation).
This source is clearly a member of an enigmatic class of X-ray
sources that are considerably more luminous than conventional X-ray binaries
but less luminous than AGN, and which are not found at the dynamical
center of the host galaxy.

The {\it Chandra} spectrum is best-fit by an absorbed power
law model with a photon index of $\Gamma = 1.8\pm{0.2}$,
similar to that seen in Galactic BH binary candidates
in their hard state. 
Bremsstrahlung models or multi-color
disk models (the favored spectral model for objects
in this class based on {\it ASCA} observations) 
can provide statistically acceptable fits only if the data at energies
$E > 5$ keV is ignored.
This is one of the first X-ray spectra of such an object 
that is unambiguously that of the source alone,
free from the spectral contamination by X-ray
emission from the rest of the galaxy that affects previous 
spectral studies of these objects using {\it ASCA}.
\end{abstract}


\section{Introduction}
\label{sec:introduction}

It has become apparent 
that there exists a class of extragalactic X-ray sources
significantly more luminous than normal X-ray binaries or supernova
remnants, but less X-ray luminous than conventional AGN such as Seyfert
nuclei (Fabbiano 1989). These intermediate-luminosity X-ray objects (IXOs,
Ptak 2001) or ultraluminous X-ray sources (ULXs, Makishima \etal (2000)) 
have X-ray luminosities in the range $10^{39}$ -- $10^{41} \ergps$ 
(equivalent to the entire Eddington luminosity for accretion onto 
a $7$ -- $700 M_{\sun}$ mass object), but
are generally {\em not} found at the dynamical center
of the galaxy (\citet{colbert1999}; \citet{roberts2000}), and hence
are not likely to be supermassive black holes (SMBHs) 
radiating inefficiently.

For example, recent {\it Chandra} High Resolution Camera (HRC)
observations of the prototypical starbursting dwarf peculiar galaxy M82
(\citet{kaaret2001}; \citet{matsumoto2001})
conclusively show that the bright X-ray variable source (peak luminosity
$L_{\rm X} \sim 9 \times 10^{40} \ergps$ in the 0.5--10.0 keV energy band) 
previously seen
in {\it Einstein}, {\it ROSAT} \& {\it ASCA} observations 
(see \citet{watson84}; \citet{bregman95} and \citet{ptak99} among
many other papers too numerous to mention)
lies $9\arcsec$ ($\sim 160 \pc$)
away from the dynamical center
of the galaxy.

The nature and origin of these objects remains unknown. The variability argues
against models of supernova remnants evolving in dense surroundings
\citep{franco1993}, where a monotonic decline in X-ray
luminosity might naively be expected (but note this is
not necessarily always true \citep{cid96}). The possible association between
the luminous X-ray source in M82 and the known young supernova remnant
41.95+57.5 \citep{stevens99}, advanced based on the source 
position derived from {\it ROSAT} HRI data, appears incorrect based on
the more accurate {\it Chandra} astrometry. No persistent counterpart to this
source at 
other wavelengths has been confirmed, although the transient
radio source 41.4+59.7 \citep{kronberg85}, seen in one observation
in 1981 but not subsequently, lies within the positional
error circle of the M82 IXO\footnote{We adopt the purely descriptive 
term IXO to refer 
to these objects over the alternative ULX as it makes no
assumptions about their nature.} \citep{kaaret2001} and may be
associated with the IXO.

The currently favored interpretation is that these might be black holes
of mass $10^{2} \la M_{\rm BH} (\Msol) \la 10^{4}$, 
intermediate between the SMBHs 
($M_{\rm BH} \sim 10^{6}$ -- $10^{9}\Msol$) found in
traditional AGN such as Seyfert galaxies (formed in some as yet unknown 
manner, which is possibly related to the growth of stellar bulges 
(\citet{gebhardt2000}; \citet{ferrarese2000})) and black holes of a few solar
masses (formed naturally as a consequence of the stellar evolution of the
most massive stars \citep{brown96}). However,
alternative explanations not involving intermediate mass black holes
do exist. \citet{king2001} argue that these X-ray sources
are conventional high-mass X-ray binaries in which the X-ray
emission happens to be beamed in our direction.
\citet{roberts2001} report a possible optical blue continuum counterpart
to an IXO in NGC 5204, which might be an extremely luminous O star
or a cluster of young stars.

Here we report on the re-detection of a strongly variable 
X-ray luminous point source 
(peak $L_{\rm X} \sim 5 \times 10^{40} \ergps$) located at least
900 pc from the nucleus of the
nearby  starburst galaxy NGC 3628 (D = 10 Mpc, Soifer \etal (1987)).
This source, discovered to be variable 
by Dahlem \etal (1995), had faded by a factor $\ga 27$ 
between two {\it ROSAT} HRI observations taken in late 1991 
and early 1994 (when it was too faint to be detected).
This source is the brightest X-ray source in a recent 
52 ks-long Chandra ACIS-S observation we obtained of NGC 3628.

With the arcsecond spatial resolution of {\it Chandra} we have been able to
measure the absolute position of this source to an accuracy of $0\farcs5$,
and obtain one of the first unambiguous X-ray
spectra of an IXO\footnote{At the initial time of writing
no other {\it Chandra} spectra of any IXO, including the M82 source, 
had been published.
This is most likely because ``pile-up'' (see \S~\ref{sec:results:spectral}) 
in the CCD detectors 
becomes a problem for point sources slightly brighter than the bright
source in NGC 3628. For sources as bright as the one in M82, it can
be so severe as to prevent any meaningful spectral analysis. 
A paper by \citet{roberts2001} appeared while this paper was undergoing the
referee process which very briefly discusses the {\it Chandra} ACIS spectrum
of an IXO in NGC 5204.}, 
free from contamination by unrelated X-ray emission 
from the rest of the galaxy.

\section{Data analysis and reduction}

NGC 3628 was observed with the {\it Chandra} ACIS-S instrument 
on 2000 December 2, for a total exposure time (after the removal
of background flares) of 52289 s. The nucleus of the galaxy
was placed at the aim point of the back-illuminated S3 chip
(the back illuminated chips are more sensitive to soft X-ray
emission, and do not suffer from the radiation damage that
compromises the spectral resolution of the front illuminated ACIS
chips). Much, but not all, of the disk of NGC 3628 falls within
the boundaries of the S3 chip.

The data was reduced and
analyzed using {\sc Ciao} (version 1.1.5),
{\sc Heasoft} (version 5.0.2) and {\sc Xspec} (version 11.0.1).
The latest {\it Chandra} calibration and spectral
response files appropriate for this
observation were used. Spectra were binned to achieve at least 20
counts per channel, to allow the use of $\chi^{2}$-fitting.
Our reduction and processing followed the guide lines released by the 
{\it Chandra} X-ray Observatory Center (CXC) and the ACIS instrument 
team\footnote{The {\sc Ciao} (v1.1.5) Science Threads
can be found at http://asc.harvard.edu/ciao1.1/threads/threads.html,
the ACIS instrument team's recipes 
at http://www.astro.psu.edu/xray/acis/recipes/.}.

The data reduction and analysis will be described in more
detail in a forthcoming paper on the diffuse X-ray emission
from this galaxy's starburst-driven superwind (Strickland \etal 2001,
in preparation).
Here we will focus only on the brightest point source within the disk
of the galaxy.

\section{Source location}
\label{sec:results:spatial}

\begin{figure}[!ht]
\plotone{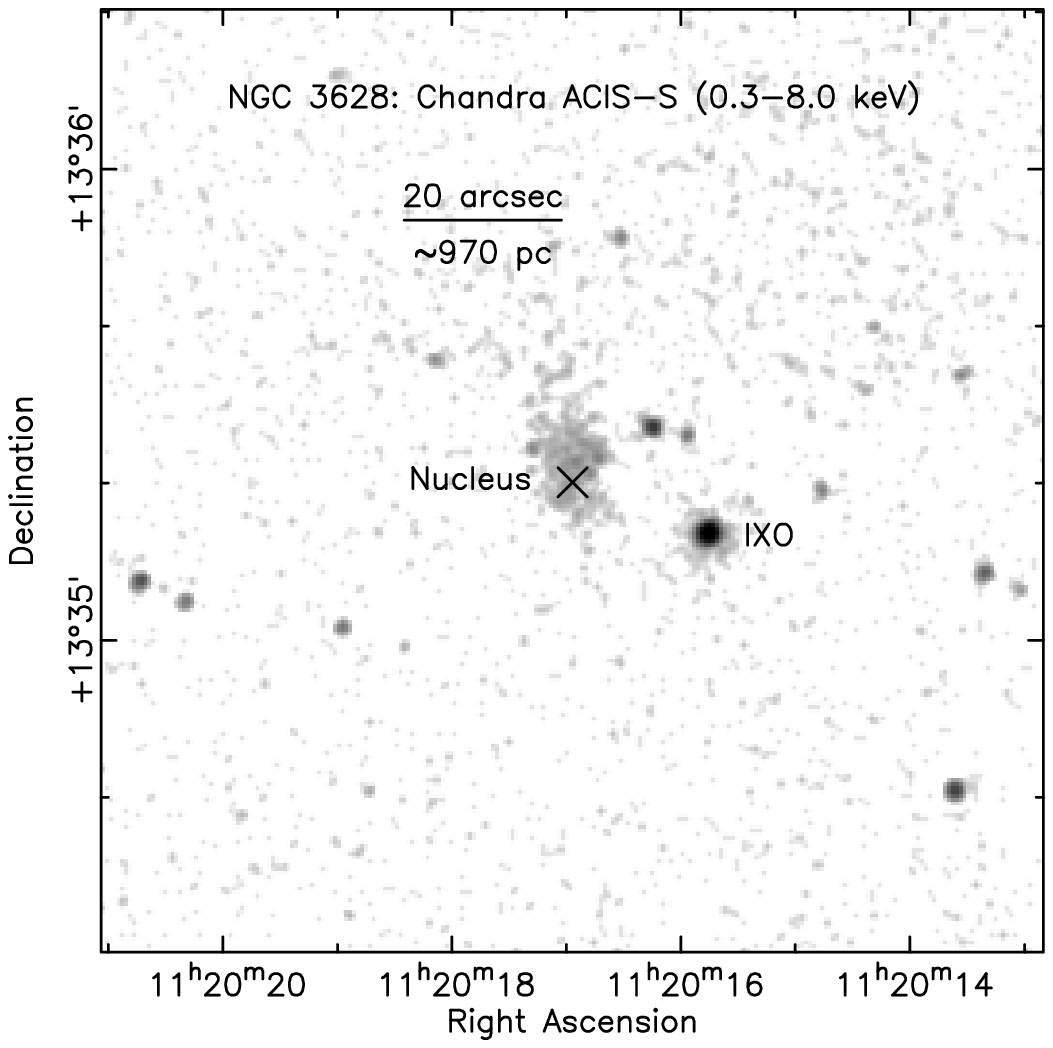}
\caption{{\it Chandra} ACIS-S3 X-ray image (0.3 -- 8.0 keV energy band) of
the central $2\arcmin$ square of NGC 3628. 
The position of the 365 MHz radio continuum peak, presumed to be the
nucleus of the galaxy, is marked by a black cross. 
The data has been smoothed with a Gaussian mask of FWHM 
$1\arcsec$, and is shown on a logarithmic intensity scale covering
3 orders of magnitude in surface brightness 
from $8\times 10^{-6}$ to $8\times 10^{-3}$ counts s$^{-1}$ arcsec$^{-2}$. 
The {\it Chandra} data also reveals wide-spread low surface
brightness diffuse X-ray emission that covers this entire region
at lower surface brightness.}
\label{fig:ximage}
\end{figure}

The {\it Chandra} observation of NGC 3628 reveals a large number of hard
X-ray point sources, embedded in highly structured and wide-spread
diffuse soft X-ray emission (see Fig.~\ref{fig:ximage}, which only
shows the central $2\arcmin$ region around the nucleus of the
galaxy). One particularly luminous point source,
at a position of $\alpha = 11^{h} 20^{m} 15\fs75$, 
$\delta = +13\degr 35\arcmin 13\farcs6$ (J2000.0),
stands out above
all the rest, and accounts for a large fraction of the hard X-ray
emission from the galaxy (see \S~\ref{sec:results:spectral}).
Following the recommended naming convention for {\it Chandra} sources,
we name this source CXOU J112015.8+133514, 
but for convenience we shall refer to it for the
remainder of this paper as the IXO.
 
Previous X-ray studies of NGC 3628 had assumed that the X-ray emission
was centered on the nucleus of the galaxy, given the limited absolute
positional accuracies of previous X-ray instruments (the highest
resolution observations prior to {\it Chandra} were with the {\it ROSAT}
 HRI, which had $1\sigma$ absolute 
positional uncertainties of $\sim 6\arcsec$).
In contrast, 
source coordinates derived from this {\it Chandra} observation are accurate
to $0\farcs5$, based on a cross-correlation between the
positions of X-ray sources in the S2 and S3 chips (excluding
X-ray sources within the disk of NGC 3628) 
and optical sources from the APM catalog \citep{maddox1990}.

With these more accurate source coordinates it is clear that the IXO is
{\em not} at the nucleus of NGC 3628, if we make the reasonable
assumption that the peaks of the K-band, H{\sc I}, CO and radio recombination
line emission correspond to the nucleus of the galaxy (See
Table~\ref{tab:nuclear_position} for references. All of these emission
maxima are within $3\arcsec$ of each other).
The nucleus of NGC 3628 is approximately $20\arcsec$ to the east of
the IXO (see Table~\ref{tab:nuclear_position}).
An offset of $20\arcsec$ corresponds
to a projected distance of 970 pc (possibly considerably 
more physically), given the assumed distance of 10 Mpc. 
Although the kinematic center
of the larger-scale H{\sc I} distribution \citep{wilding93} is offset 
to the north-west of the nucleus by about $12\arcsec$, 
it is also $\sim 20\arcsec$ from the IXO.

As this source has a luminosity only a factor $\sim 4$ lower than the
peak luminosity of the 
bright source seen by the {\it ROSAT} HRI 
(see \S~\ref{sec:results:variability}), and there were no other
comparable luminosity sources seen in previous observations,
it seems most probable that the IXO is the same source
as the variable HRI source discussed in \citet{dahlem95}.
Comparison of the {\it ROSAT} HRI images with the 0.3 -- 2.0 keV
{\it Chandra} images supports this interpretation, based on the
relative position of other weak 
X-ray sources found in both sets of observations.

We have searched for an optical, near-IR or radio counterpart
to this source. R-band optical images \citep{fhk90} reveal the
location of the IXO to be within one of the 
dust lanes obscuring much of the disk of NGC 3628. Within
a $2\arcsec$ radius of the {\it Chandra}-derived position of the IXO
the optical surface brightness is relatively uniform, and there
are no obvious star clusters or other features visible.
K-band imaging, taken by Colbert, also revealed no features
within a $5\arcsec$ radius of the IXO's position.
No radio source was detected in the 15~GHz VLA A-configuration data from
\citet{carral90}.  The 3$\sigma$ upper limit of 0.24 mJ/beam 
corresponds to
a radio power P $<$ 3 $\times$ 10$^{18}$ W~Hz$^{-1}$, about an
order of magnitude more luminous than the galactic supernova
remnant Cas~A, but $\sim 10$ -- 1000 
times lower than typical radio supernovae \citep{weiler86}.

\section{Spectral properties and count rates}
\label{sec:results:spectral}

\subsection{Spectral fitting}

\begin{figure}[!ht]
\epsscale{0.70}
\plotone{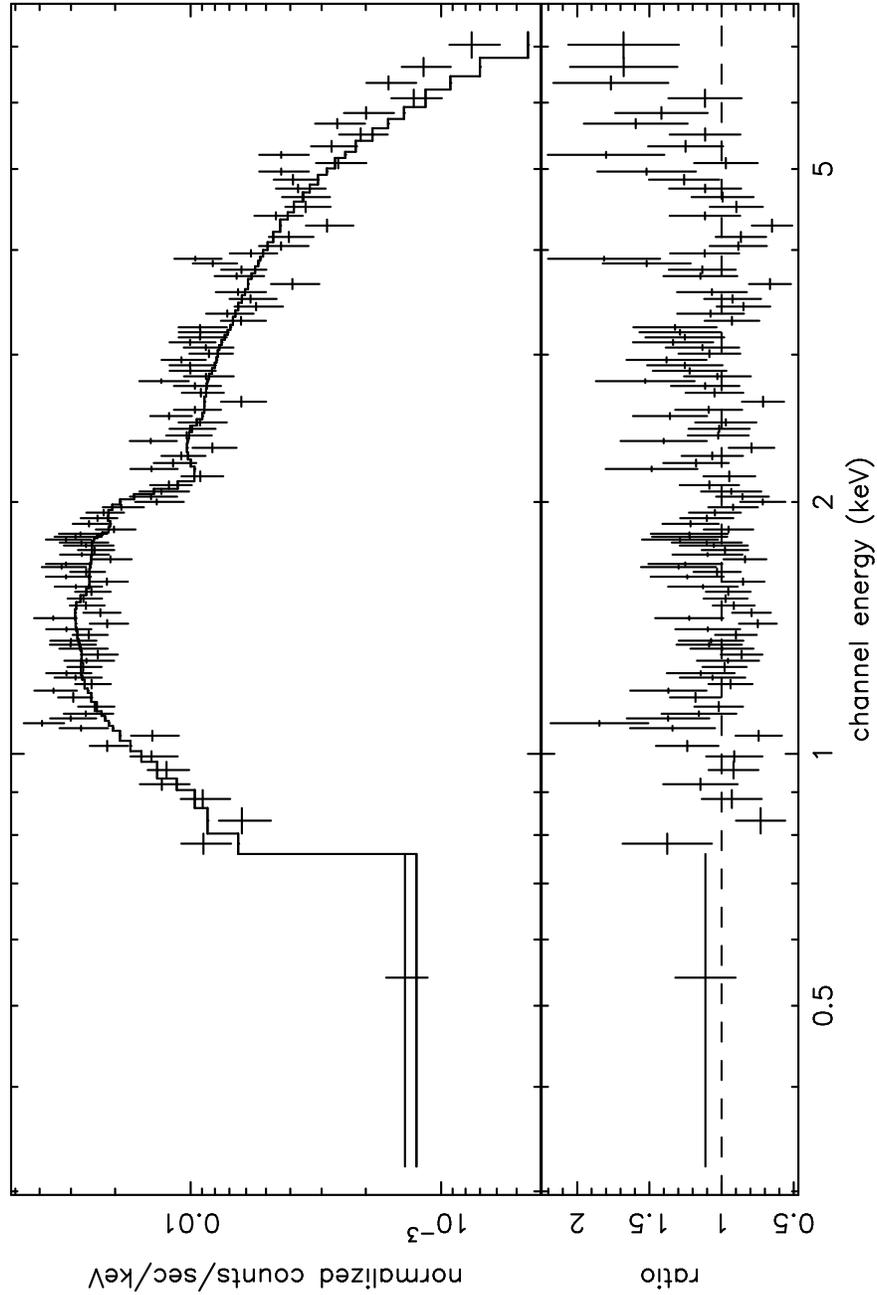}
\caption{Chandra ACIS-S3 spectrum of the IXO. The solid line shows
 the best-fitting absorbed power law model fit to the data in the
 energy range 0.3 -- 8.0 keV (see \S~\ref{sec:results:spectral} 
 and Table~\ref{tab:specfits}). 
 The lower panel shows the
 ratio of the data to the model.}
\label{fig:spectrum}
\end{figure}

Fig.~\ref{fig:spectrum} shows the ACIS spectrum of the IXO,
extracted using a circular aperture of radius $6\arcsec$. 
For the purposes of background subtraction
 we used a circular aperture of radius $30\arcsec$,
offset $20\arcsec$ directly to the west of the IXO, having
excluded the region around the bright source itself and around the
five other {\it Chandra}-detected X-ray sources falling within this region.
Use of a local background region is warranted, given the presence of
extended diffuse soft X-ray emission (associated with
the starburst-driven wind) throughout the disk of NGC 3628.

We fit a variety of different spectral models to the {\it Chandra} spectrum
using the data in the energy range 0.3 -- 8.0 keV,
focusing mainly on models 
traditionally associated with AGN (absorbed power laws) or
other IXOs observed with ASCA (multi-color disk models,
see \citet{makishima2000}). 

An absorbed power law
provides the simplest statistically acceptable fit to the spectrum
(Fig.~\ref{fig:spectrum}, table~\ref{tab:specfits}).
Absorbed multi-color disk, bremsstrahlung or {\sc Mekal} hot
plasma models \citep{mewe95} are all statistically unacceptable
fits to the IXO spectrum (reduced $\chi^{2} \sim 1.1$ -- $1.4$).

Statistically good fits can be obtained using these models 
if an additional hard component was added to the model.
Two component spectra, comprising a soft multi-color disk with a
hard power law tail, have been found for other IXOs using
{\it ASCA} spectroscopy (\citet{takano94}; \citet{colbert1999}). 
The model parameters of the added power law or Gaussian line model
used as this hard component are very poorly constrained, and physically some
of the parameters are not realistic, so it is not clear if this
apparent hard component is real or not.


A small fraction of the events detected from this source will be
due to pile-up (the mis-identification of two or more X-ray photons 
falling within the same pixel during a single CCD exposure of 
3.2s as a single photon of higher energy). This has many affects,
the simplest of which is a systematic hardening of the spectrum,
which is a possible cause of
the additional hard component the spectral fits appear to require.
The {\sc Pimms} count-rate calculator predicts $\sim 8$\% of the 
total counts are piled-up given the observed count rate of
this source. This is close to the 10\% level at which 
the CXC warn that spectra are likely to be significantly affected,
suggesting  pile-up may be a problem. 

To investigate whether there is appreciable pile-up in the NGC 3628 IXO
spectrum we compared the event grade branching ratios (the fraction
of events with a given grade, see the {\it Chandra} 
Proposers Observatory Guide) for photons
with $2 \le E$ (keV) $\le 8$ from the IXO (1646 events)
with the summed emission from the 14 next brightest point sources 
 seen in NGC 3628 which are individually too
faint to suffer from any pile-up (672 events in total). Pile-up leads to a 
systematic increase in the fraction of events with high grades
(\eg ASCA grades 5, 6 \& 7). However, we find that the event grade branching
ratios for the IXO and the summed fainter point source emission are
identical to within $2\sigma$ for all grades, which suggests that there is
no appreciable pile-up in the NGC 3628 IXO spectrum. Applying this
technique to the public 34 ks {\it Chandra} ACIS-I observation of M82
(ObsID 361) shows a very significant migration of the events from the
M82 IXO to higher grades with respect to the fainter point sources in M82.

Other evidence that the hard excess seen in the NGC 3628 IXO is
not due to pile-up is that
spectral fitting of a brighter point source seen in our 
two {\it Chandra} observations
of NGC 253 (Weaver \etal, in preparation) shows no sign of a
hard excess, even though the predicted pile-up fraction is
12 -- 14\% (higher than the NGC 3628 IXO), 
and the total number of counts in the spectrum
is similar to that in the NGC 3628 IXO spectrum.

Statistically acceptable multi-color disk spectral models
{\em can} be obtained if we restrict the energy range used to 0.3 -- 5.0 keV
(\ie deliberately
excluding data in the energy range where the multi-color disk 
model does worst).
However, it should be noted that 
absorbed power law and absorbed bremsstrahlung models 
provide marginally better fits to
the data in this restricted energy range.
It therefore seems that a power law model provides the best description
of the current spectral state of this source, and that a multi-color
disk model does {\em not} provide an adequate fit to the spectrum of the
IXO.

Source count rates and absorption-corrected fluxes and luminosities,
assuming the power law spectral model (using the 
restricted 0.3 -- 5.0 keV energy range), are given in Table~\ref{tab:fluxes}.
All spectral fits give absorption columns in the range $\nH = 5$ --
$8 \times 10^{21} \pcmsq$. These columns are consistent with foreground
absorption of the source by the edge-on disk of NGC 3628, and its
location in or behind the optical dust lane. 

\subsection{Comparison to previous spectral studies of NGC 3628}
It is not easy to compare our {\it Chandra} spectra of the IXO
with previous X-ray spectral studies of the nucleus NGC 3628,
using the {\it ROSAT} PSPC \citep{dahlem95} or {\it ASCA} \citep{yaqoob95},
given the significant variability of the IXO and the much poorer spatial
resolution of the earlier observations (The {\it ROSAT} PSPC spectrum covers a
region $2\arcmin = 5.8$ kpc in radius, and the {\it ASCA} spectra a 
region $3\arcmin = 8.7$ kpc in radius). 

The most detailed existing spectral study is that of \citet{dwh98},
who performed a joint spectral fit of the PSPC and {\it ASCA}
spectra, taking into account the variability between the two observations.
This joint spectrum is best characterized by two soft thermal
components and a hard absorbed power law. The photon index and
hydrogen column derived for the power law component 
($\Gamma = 1.63^{+0.14}_{-0.17}$ and $\nH = 9\pm{2} \times 10^{21} \pcmsq$) 
agree well with the {\it Chandra}-based power law fit to the emission 
from the IXO.

The close agreement between the power law slopes may be fortuitous,
as most of the hard X-ray counts in the ASCA spectrum must have been
from other point sources in NGC 3628 and not from the IXO itself.
Even in its present state, which is several times more luminous than
it was during the 1993 ASCA observations 
(see \S~\ref{sec:results:variability}), hard X-ray emission from sources
other than the IXO still provide a significant fraction ($\sim 33$\%)
of the 2.0 -- 8.0 keV count rate in the central region of NGC 3628.
Within a $3\arcmin$ radius region, the
{\it Chandra} observations reveal at least 33 other
point-like X-ray sources, in addition to the diffuse thermal emission
associated with the starburst. We find that the IXO only 
provides $\sim35$\%
of the ACIS-S3 0.3 -- 8.0 keV energy band count rate within
a radius of $3\arcmin$ from the source. In the 2.0 -- 8.0 keV
energy band the IXO is more dominant, providing
67\% of the ACIS-S3 counts, {\em but emission from the rest of the
galaxy is always significant}.

We estimate that these other point sources, along with the diffuse emission,
account for $\sim 80\pm{30}$\% of the total {\it ASCA} count rate
(the main uncertainty is the uncertainty in the {\it ASCA} count rate
\citep{yaqoob95}). As the IXO, at the luminosity observed in December 2000,
would produce $\sim90$\% of the total {\it ASCA} count rate itself,
the IXO must have been less luminous in 1993 than it is now.
The shape of the 
hard spectral component in the joint {\it ROSAT} PSPC and {\it ASCA}
spectral fit was therefore largely determined by the
30 or so other hard X-ray point sources within the 
central $3\arcmin$ of NGC 3628.

It is worth noting that 
{\it Chandra} provides {\em a spectrum that 
is unambiguously that of the IXO alone} 
--- the same can not be so easily said for 
{\it ASCA} spectra of IXOs, which are typically of a
region a few arcminutes in radius.
The luminosity of these objects
is very similar to that of the rest of the normal spiral or starburst galaxies
they inhabit. That contamination of the {\it ASCA} spectra by unrelated binary
and diffuse X-ray emission may be biasing
{\it ASCA} spectral fits in favor of the multi-color disk spectral
model is a legitimate concern.

To investigate whether spectral contamination (at the level of 
$\sim 30$\% of the hard X-ray counts that we find in NGC 3628)
does bias the fitted spectral shape of the hard X-ray emission, we extracted
a spectrum of all the emission within a 3 arcminute radius of the IXO
(approximately the same region an {\it ASCA} spectrum would cover).
We fit the spectrum with spectral models comprising two absorbed
soft thermal components and an absorbed hard component (either a
power law or a multi-color disk model), based on the good empirical
fit such three component models give to the combined {\it ROSAT} PSPC and
{\it ASCA} spectra of starbursts \citep{dwh98}. For the {\it Chandra}
spectrum, two soft thermal components are required to obtain a 
statistically acceptable fit, but the hard component can equally
well be fitted by a multi-color disk or a power law (the difference
between the two fits is not significant: $\Delta \chi^{2} = 0.7$ for
a change by 1 in the number of degrees of freedom). The
spectral shape of the fitted hard component (best-fitting
 \nH, $\Gamma$ or $kT$)
was statistically consistent with the spectral fits to the IXO alone,
except for a higher model normalization due to the flux from the
33 other X-ray point sources.

This suggests that the reason we do not find the IXO spectrum
to be clearly better fit by the multi-color disk model strongly
favored by {\it ASCA} observations of other IXOs 
(\citet{colbert1999}; \citet{makishima2000})
is due to a real spectral difference between this source and the
other IXOs, and not due to spectral contamination (at least not
spectral contamination at the 30\% level).
One possibility is that IXOs can change spectral state, from
soft states described by the multi-color disk model to harder
power law spectral states (see \citet{colbert1999} and \citet{kubota2000}), 
and that we are observing the NGC 3628
IXO in such a hard spectral state.

\section{Source variability}
\label{sec:results:variability}

This source displays strong variability on the time scale of
months or years (Dahlem \etal 1995). We are now able
to improve previous estimates of the X-ray flux of this source
over time by making use of the better-determined {\it Chandra} spectrum,
and by making use of {\it Chandra}'s spatial resolution to correct
the {\it Einstein}, {\it ROSAT} PSPC and {\it ASCA} flux
measurements for the contribution from the other point sources and diffuse
emission.

Within the $\sim 17$ hour duration of our { \it Chandra} observation
the count rate of the bright source, in either the broad 
(0.3 -- 8.0 keV) or hard energy band (2.0 -- 8.0 keV), is 
statistically consistent with being constant (although this does
not exclude genuine variation on the order of $\sim 10$\% or less).

\begin{figure}[!ht]
\plotone{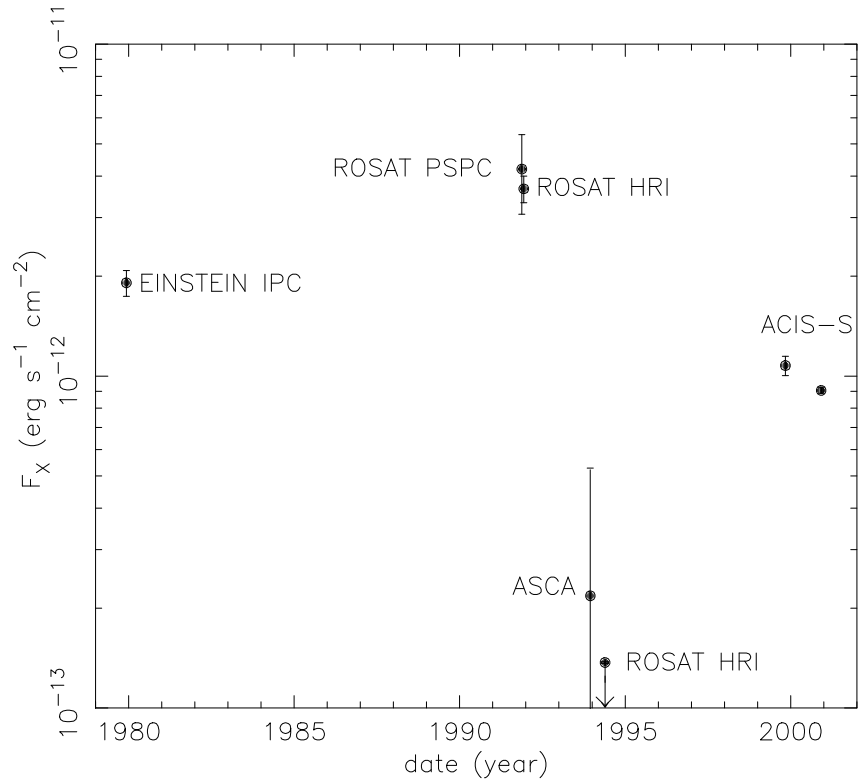}
\caption{Our best estimate of the 
        X-ray flux (in the 0.3 -- 8.0 keV energy band) of the NGC 3628 IXO
	from 1979 Dec. through to 2000 Dec.,
	based on the observed count rates from a variety of X-ray
	instruments, assuming
  	the {\it Chandra}-observed spectral shape has remained constant.
  	The lower-spatial resolution 
	{\it Einstein}, {\it ROSAT} PSPC and {\it ASCA} measurements
	have been corrected to take account of unrelated point source
	and diffuse X-ray emission falling within the spectral-extraction
	region. Error bars represent $1\sigma$
	uncertainties in observed count rates alone.}
\label{fig:historical}
\end{figure}

Fig.~\ref{fig:historical} presents our best estimate of the
absorption-corrected X-ray flux of this object with time, 
based on these observations, the papers of
\citet{dahlem95}, \citet{fhk90}, \citet{dahlem96} \& 
\citet{yaqoob95}, and
the count rate in a 2 ks {\it Chandra} ACIS-S3 observation taken 
on 1999 November 3, as part of a {\it Chandra} GTO program 
(Ptak 2000, private communication). 

We used the power law model that best-fits the Chandra spectrum to 
predict {\it Einstein}, {\it ROSAT} (PSPC \& HRI) and 
{\it ASCA} count rates. The ratio of the observed to
predicted count rate, for any particular observation, multiplied
by the absorption corrected X-ray flux
(0.3 -- 8.0 keV energy band) from this Chandra observation
is the estimated X-ray flux plotted in Fig.~\ref{fig:historical}. 

For the {\it Einstein}, {\it ROSAT} PSPC and
{\it ASCA} observations we have calculated a correction to obtain
the flux due to the bright source alone (given the significant flux
from other point sources and diffuse emission, see 
\S~\ref{sec:results:spectral}). 
We estimate that approximately
48\% of the {\it Einstein} count rate, 33\% of the
{\it ROSAT} PSPC count rate and 78\% of the {\it ASCA} count rate
was due to diffuse X-ray emission and point sources other than the IXO.

The relative fluxes of the IXO during the Einstein ROSAT and ASCA
observations differ somewhat from the work of Dahlem \etal (1995) 
and \citet{dwh98}, due to
our correction for emission from other point sources and diffuse gas,
and due to the strong dependence of absorption-corrected
flux on the assumed spectral shape.

The IXO was substantially brighter in the early 1990s than it
is now, with an intrinsic X-ray luminosity (0.3 -- 8.0 keV energy band)
of $\sim 5 \times 10^{40} \ergps$. This makes it one of the most
luminous IXOs, only slightly less luminous than the source in
M82 (Kaaret \etal 2001; Makishima \etal 2000). A drop in flux by a factor 
$\gtrsim 27$ occurred between late 1991 Dec. and 1994 May (Dahlem \etal 1995).
Our estimate of a low X-ray flux from the IXO in the 1993 December {\it ASCA}
observation is consistent with this fading, 
given the non-detection of the source by the 
{\it ROSAT} HRI only 5 months later.

One should bear in mind that Fig.~\ref{fig:historical} is based on 
the assumption that spectral shape of the source has remained the
same, only changing in absolute luminosity. Given the large variations
in derived luminosity, it is quite possible that the spectral shape
has changed, due to intrinsic changes in the source itself or possibly
even variations in foreground absorption (as discussed by \citet{dahlem95}).

\section{Discussion, speculation and conclusions}
\label{sec:discussion}

What is this X-ray-luminous source?
Given the substantial offset of the brightest X-ray source from the nucleus of
NGC 3628 ($\sim 20\arcsec \equiv 970$ pc), this object can not be
a SMBH inefficiently radiating X-rays, as dynamical
friction would make a SMBH fall to the center of the galaxy within
an astronomically short time. This object is clearly another example of
an IXO (or ULX).  Is it a candidate intermediate
mass black hole, or can it be explained by models involving X-ray-luminous
supernovae or beamed X-ray emission from normal BH binaries?

Interpretations of this source as either a young X-ray 
luminous SNR or a supernova remnant 
evolving in a high-density medium \citep{franco1993} have problems.
Most young SNRs have soft thermal X-ray spectra \citep{schlegel95},
unlike the observed {\it Chandra} spectrum of this object 
(\S~\ref{sec:results:spectral}).
Spectrally this source is consistent with some
models of supernovae in high-density environments which predict
power-law-like X-ray spectra with photon indices of $\Gamma = 1.6$
-- $2.0$ \citep{plewa95}, as the power law spectral model
fit to the Chandra spectrum gives  $\Gamma = 1.8\pm{0.2}$. 
Nevertheless, the pattern of source 
variability seen (the observed increase in flux after the original
peak and strong decline, see \S~\ref{sec:results:variability}), 
along with an upper limit on the radio flux at this location
significantly lower than known radio supernovae 
(\S~\ref{sec:results:spatial}), are convincing evidence against
a supernova-related interpretation.

Although the {\it ASCA} spectra of most other known IXOs are clearly
better fit by a multi-color disk model than a power law model
(\citet{colbert1999}; \citet{makishima2000}),
our {\it Chandra} spectrum is  better fit by a power
law than by a multi-color disk model. 
Three IXOs observed with {\it ASCA} appear to show
changes in spectral state, from a soft state well
represented by a multi-color disk to a hard state best
modeled as a power law (\citet{colbert1999}; \citet{kubota2000}),
and it is possible that the NGC 3628 source represents another
IXO in such a hard spectral state.
Our fitted power law slope, $\Gamma = 1.8\pm{0.2}$,
is very similar to that seen in Galactic BH binary candidates
in their hard state (see references in \citet{esin98}),
or to the canonical $\Gamma = 1.7$ power law of AGN.
More {\it Chandra} spectra of
IXOs will be necessary to convincingly determine
the spectral properties of this class of object.


With an absorption-corrected 0.3 -- 8.0 keV X-ray luminosity
of $1.1 \times 10^{40} \ergps$ in this observation (assuming the
emission to be isotropic), and a peak
luminosity of perhaps $\sim 5 \times 10^{40} \ergps$ in the early
1990s, this object is one of the more luminous IXOs known (others
of this class more typically have 0.5 -- 10 keV luminosities
of $10^{39}$ -- $10^{40} \ergps$ (Makishima \etal 2000), while the
luminous source in M82 reaches $\sim 10^{41} \ergps$ at maximum).
Both of these objects display strong variability, with
X-ray luminosities changing by at least an order-of-magnitude
over a period of a few years.

This object is not the only IXO in NGC 3628. Another luminous
X-ray source $\sim 5\arcmin$ to the east of the nucleus (but
still within the disk of NGC 3628) was detected in the earlier {\it ROSAT}
and {\it ASCA} observations (source 14 in \citet{dahlem96}, see also
\citet{yaqoob95}). Unfortunately this source lies outside the
region covered in these Chandra observations, but should be covered
in the recent (currently unpublished) {\it XMM-Newton} observation
of NGC 3628.
Based on the relative {\it ROSAT} PSPC count
rates and assuming a similar spectrum for both objects, the
second IXO had an absorption-corrected 0.3 -- 8.0 keV X-ray luminosity
of $1.2 \times 10^{40} \ergps$ in late 1991.

That both M82 and NGC 3628 are also both archetypal starburst galaxies
is perhaps a coincidence, but it is interesting to speculate
on the possibility of a link between the formation of large
numbers of massive stars and the growth of massive black holes
from stellar mass remnants (Taniguchi \etal 2000). 
There is some form of relationship between
starbursts and type 2 Seyfert galaxies (Heckman \etal 1989;
\citet{terlevich90a}; \citet{terlevich90b};
Cid Fernandes \& Terlevitch 1995; Gonz\'{a}lez Delgado,
Heckman \& Leitherer 2001; 
Levenson, Weaver \& Heckman 2001a, 2001b), although not necessarily
directly evolutionary in kind. If IXOs are intermediate mass
black holes, might they be a ``missing-link''
connecting star formation and black hole formation?

The model of IXOs as relatively normal BH X-ray binaries
(or perhaps analogues of the Galactic microquasars) remains
a strong contender. As \citet{king2001} argue, the apparently
high numbers of IXOs in star-forming galaxies may be a problem
for models invoking intermediate mass black holes. In contrast,
elevated chances of seeing a beamed (but otherwise normal) X-ray binary
are expected in starburst galaxies. The high-amplitude variability
seen in both the NGC 3628 and M82 IXOs is also a natural
consequence of this model (see \citet{king2001}).

Testing the various hypotheses currently in the literature requires
coordinates of IXOs accurate to $\la 1\arcsec$,
in order to search for counterparts at other wavelengths
(\eg very dense young stellar clusters, or binary companions 
under the beaming hypothesis)  and to
study the environments in which these objects are found and may
have formed in. This is a job only {\it Chandra} can perform.

It is clear that {\it Chandra} has much to offer in the study of these
X-ray sources. As we have demonstrated with this
study of the IXO in NGC 3628, it is possible to obtain 
absolute positions accurate to within
$0\farcs5$, and spectra free from contamination by unrelated
X-ray point sources and diffuse emission.
Both of these capabilities are vital to further the study of these
intriguing objects.

\acknowledgments

It is a pleasure to thank the anonymous referee for their helpful
comments.
We would like to thank Andy Ptak for kindly providing us with the count
rate of the IXO from the short 1999 November 3rd {\it Chandra} observation.
EJMC thanks Patricia Carral for use of her calibrated VLA data.
DKS is supported by NASA through {\it Chandra} Postdoctoral Fellowship Award
Number PF0-10012, issued by the {\it Chandra} X-ray Observatory Center,
which is operated by the Smithsonian Astrophysical Observatory for and on
behalf of NASA under contract NAS8-39073.



\newpage

\newpage

\newpage

\newpage

\begin{deluxetable}{lcclr}
\tabletypesize{\scriptsize}
\tablecolumns{4}
\tablewidth{0pc}
\tablecaption{The location of the nucleus of NGC 3628 
	and the IXO \label{tab:nuclear_position}}
\tablehead{
\colhead{Observed feature} 
	& \colhead{$\alpha$ (J2000)} 
	& \colhead{$\delta$ (J2000)} 
	& \colhead{Reference} & \colhead{Offset\tablenotemark{a}} \\
\colhead{}
	& \colhead{(h m s)} 
	& \colhead{($\degr$ $\arcmin$ $\arcsec$)}
	& \colhead{} & \colhead{$\arcsec$}
}
\startdata
Peak of central H{\sc I} disk 
	& 11 20 17.03 & +13 35 19.95 & \citet{cole98}  & 19.6 \\
Radio recombination line peak 
	& 11 20 17.02 & +13 35 19.85 & \citet{zhao97}  & 19.4 \\
CO molecular disk 
	& 11 20 17.00 & +13 35 19.85 & \citet{irwin96} & 19.2 \\
K-band nucleus     
	& 11 20 17.00 & +13 35 22.60 & Unpublished imaging by Colbert & 20.3 \\
365 Mhz peak 
	& 11 20 16.95 & +13 35 20.10 & \citet{douglas96} & 18.6 \\
Large-scale H{\sc I} kinematic center
	& 11 20 16.46 & +13 35 29.60 & \citet{wilding93} & 18.9 \\
Secondary peak in CO emission
	& 11 20 15.96 & +13 35 20.90 & \citet{irwin96}   & 7.9\\
\tableline
IXO (CXOU J112015.8+133514)
	& 11 20 15.75 & +13 35 13.60 & This work & \nodata \\ 
\enddata
\tablenotetext{a}{Angular distance 
	from this position to the IXO in arcseconds.}
\end{deluxetable}

\newpage

\begin{deluxetable}{llllllllll}
 \tabletypesize{\scriptsize}%
\tablecolumns{10} 
\tablewidth{0pc} 
\tablecaption{Best-fitting spectral models for the IXO Chandra spectrum
	\label{tab:specfits}} 
\tablehead{ 

\colhead{XSPEC model\tablenotemark{a}} & 
	\colhead{$N_{\rm H}$\tablenotemark{b}} &
	\colhead{$\Gamma$\tablenotemark{c}} & 
	\colhead{Norm.\tablenotemark{d}} &
	\colhead{$kT$\tablenotemark{e}} & \colhead{Norm.\tablenotemark{f}} &
	\colhead{$EW_{\rm Fe-K}$\tablenotemark{g}} & 
	\colhead{$\sigma$\tablenotemark{h}} & 
	\colhead{$\chi^{2}_{\nu}$} 
	& \colhead{$\chi^{2}$ ($\nu$)\tablenotemark{i}}
}
\startdata
\cutinhead{Statistically acceptable fits with 
	well-constrained model parameters}
wabs*pow & $7.5^{+1.1}_{-0.8}$
	& $1.69^{+0.15}_{-0.14}$ & $1.44^{+0.33}_{-0.25}$
	& \nodata & \nodata
	& \nodata & \nodata
	& 1.03 & 111.4 (108) \\
{\bf wabs*pow\tablenotemark{j}} & $8.0^{+1.2}_{-1.1}$
	& $1.82^{+0.21}_{-0.20}$ & $1.63^{+0.41}_{-0.32}$
	& \nodata & \nodata
	& \nodata & \nodata
	& {\bf 0.97} & {\bf 93.9 (97)} \\
wabs*(pow+gauss) & $7.6^{+1.2}_{-1.1}$
	& $1.71^{+0.19}_{-0.19}$ & $1.47^{+0.38}_{-0.29}$
	& \nodata & \nodata
	& $0.16^{+0.27}_{-0.16}$ & $0.01$ (f)
	& 1.01 & 108.3 (107) \\
wabs*bremms\tablenotemark{j} & $7.1^{+1.0}_{-0.8}$
	& \nodata & \nodata
	& $5.9^{+3.5}_{-1.7}$ & $1.74^{+0.26}_{-0.19}$
	& \nodata & \nodata
	& 0.96 & 93.1 (97) \\
wabs*diskbb\tablenotemark{j} & $5.8^{+0.8}_{-0.7}$
	& \nodata & \nodata
	& $1.38^{+0.20}_{-0.17}$ & $8.67^{+5.40}_{-3.48}$
	& \nodata & \nodata
	& 1.01 & 97.7 (97) \\
\cutinhead{Statistically unacceptable fits, or models with 
	poorly constrained physical parameters}
wabs*brems & $6.6^{+0.9}_{-0.6}$
	& \nodata & \nodata
	& $9.4^{+5.8}_{-2.7}$ & $1.62^{+0.17}_{-0.11}$
	& \nodata & \nodata
	& 1.09 & 117.8 (108) \\
wabs*mekal & $6.4^{+0.7}_{-0.6}$
	& \nodata & \nodata
	& $10.3^{+5.7}_{-2.6}$ & $4.33^{+0.27}_{-0.26}$
	& \nodata & \nodata
	& 1.14 & 123.1 (108) \\
wabs*mekal\tablenotemark{j} & $6.8^{+0.8}_{-0.7}$
	& \nodata & \nodata
	& $7.1^{+3.6}_{-1.7}$ & $4.36^{+0.34}_{-0.30}$
	& \nodata & \nodata
	& 1.07 & 104.2 (97) \\
wabs*diskbb & $5.2^{+0.7}_{-0.5}$
	& \nodata & \nodata
	& $1.71^{+0.26}_{-0.21}$ & $4.05^{+2.44}_{-1.61}$
	& \nodata & \nodata
	& 1.35 & 146.3 (108) \\
wabs*(pow+gauss) & $8.0^{+2.0}_{-1.3}$
	& $1.83^{+0.35}_{-0.25}$ & $1.64^{+0.86}_{-0.39}$
	& \nodata & \nodata
	& $1.3^{+14.3}_{-1.3}$ & $0.9^{+19.1}_{-0.9}$
	& 0.96 & 102.1 (106) \\
wabs*(diskbb+gauss) & $5.3^{+0.8}_{-0.7}$
	& \nodata & \nodata
	& $1.66^{+0.29}_{-0.22}$ & $4.43^{+3.05}_{-1.90}$
	& $0.37^{+0.39}_{-0.37}$ & $0.01$ (f)
	& 1.30 & 139.3 (107) \\
wabs*(diskbb+gauss) & $6.1^{+1.6}_{-1.0}$
	& \nodata & \nodata
	& $1.19^{+0.35}_{-0.49}$ & $14.3^{+78.2}_{-8.7}$
	& $12.5^{+116.8}_{-7.8}$ & $1.45^{+18.55}_{-0.78}$
	& 0.96 & 101.8 (106) \\
wabs*(diskbb+pow) & $6.3^{+2.5}_{-1.1}$
	& $-0.7^{+3.7}_{-2.3}$ & $0.02^{+1.46}_{-0.02}$
	& $0.99^{+0.46}_{-0.99}$ & $26.0^{+806.6}_{-19.2}$
	& \nodata & \nodata
	& 0.96 & 102.1 (106) \\
\enddata 

\tablenotetext{a}{Model specification within {\sc Xspec}. Model components
	used were {\it wabs} (foreground X-ray absorption), {\it pow} (power
	law model), {\it gauss} (a Gaussian line), {\it diskbb} (multi-color
	disk model), 
	{\it brems} (thermal bremsstrahlung) \& {\it mekal} ({\sc Mekal} hot
	plasma model with Solar element abundances).}
\tablenotetext{b}{Hydrogen column, in units of $10^{21} \pcmsq$.}
\tablenotetext{c}{Power law model photon index.}
\tablenotetext{d}{Power law model normalization, in units of 
	$10^{-4}$ photons keV$^{-1}$ $\pcmsq \ps$ at 1 keV.}
\tablenotetext{e}{Characteristic temperature in keV. For the multi-color disk 
	model this is temperature at the inner disk radius, for the 
	bremsstrahlung and hot plasma models this is the gas temperature.}
\tablenotetext{f}{Model normalization. For the multi-color disk model this
	is in units of $10^{-3} (R_{\rm in}/D_{\rm 10 kpc})^{2} \cos \theta$,
	where  $R_{\rm in}$ is the inner radius of the disk in km,
        $D_{\rm 10 kpc}$ is the distance to the source in units of 10 kpc, 
	and $\theta$ the inclination of the disk with respect to the 
	line of sight. For the 	bremsstrahlung and hot plasma models the
	normalization is in units of $10^{-4} K$. For the thermal 
	bremsstrahlung model $K = 3.02 \times 10^{-15} \int n_{\rm e} 
	n_{\rm i} dV / 4\pi D^{2}$, where $n_{\rm e}$ and $n_{\rm i}$
	are the electron and ion number densities in units of cm$^{-3}$,
	$dV$ a volume element and $D$ the distance to the source in cm.
	For the {\sc Mekal} hot plasma model, $K = 10^{-14} \int n_{\rm e} 
	n_{\rm H} dV / 4\pi D^{2}$, where $n_{\rm H}$ is the proton number
	density in units of cm$^{-3}$.}
\tablenotetext{g}{Equivalent width of the the Gaussian line component,
	with a fixed central energy of 6.4 keV.}
\tablenotetext{h}{Width ($1\sigma$) of the 6.4 keV Gaussian line feature 
	in keV.}
\tablenotetext{i}{Fit $\chi^{2}$ and number of degrees of freedom ($\nu$).}
\tablenotetext{j}{Fits using only the data in the energy band 0.3 -- 5.0 keV.}

\tablecomments{Unless otherwise noted all fits are to data 
	in the 0.3 -- 8.0 keV energy band. 
	All errors quoted are 90\% confidence for a number
	of interesting parameters equal to the number of free parameters
	in the fit. Parameters fixed during fitting are denoted by (f).
	Our preferred spectral model (an absorbed power law) has been
	highlighted in bold face type.}
\end{deluxetable} 

\newpage

\begin{deluxetable}{lccc}
\tabletypesize{\scriptsize}
\tablecolumns{4}
\tablewidth{0pc}
\tablecaption{IXO count rates and X-ray fluxes \label{tab:fluxes}}
\tablehead{
\colhead{Energy\tablenotemark{a}} & \colhead{Rate\tablenotemark{b}}
	& \colhead{$F_{\rm X}$\tablenotemark{c}} 
	& \colhead{$L_{\rm X}$\tablenotemark{d}}
}
\startdata
0.3 -- 8.0 & $5.36\pm{0.10}$ & 9.40 & $1.1 \times 10^{40}$ \\
0.3 -- 2.0 & $2.83\pm{0.07}$ & 4.75 & $5.7 \times 10^{39}$ \\
2.0 -- 8.0 & $2.53\pm{0.07}$ & 4.65 & $5.6 \times 10^{39}$ \\
\enddata
\tablenotetext{a}{Energy range in keV.}
\tablenotetext{b}{Background-subtracted ACIS-S3 count rate, in units of
	$10^{-2}$ count $\ps$.}
\tablenotetext{c}{Absorption-corrected (\ie intrinsic) X-ray flux, based
	on the best-fitting absorbed power law model (fit only within the
	0.3 -- 5.0 keV energy range), in units of $10^{-13}$ erg 
	$\ps \pcmsq$. Derived fluxes depend on which spectral model
	is used, and vary from these values by up to $^{+10}_{-30}$\%
	if other spectral models from Table~\ref{tab:specfits} are used.}
\tablenotetext{d}{Absorption corrected X-ray luminosity in units of 
	$\ergps$, assuming D = 10 Mpc.}
\end{deluxetable}

\end{document}